\documentclass[amsmath,amssymb]{revtex4-2}
\usepackage{graphicx}
\usepackage{subfig}
\usepackage{epsfig}
\usepackage{dcolumn}
\usepackage{multirow}
\usepackage{rotating}
\usepackage{verbatim}
\usepackage{bm}
\usepackage{color}
\usepackage{soul}
\usepackage{float}
\usepackage[utf8]{inputenc}
\usepackage{physics}
\usepackage[colorlinks=true, linkcolor=blue, citecolor=blue, urlcolor=blue]{hyperref}
\def\lsim{\raise0.3ex\hbox{$<$\kern-0.75em\raise-1.1ex\hbox{$\sim$}}}
\def\gsim{\raise0.3ex\hbox{$>$\kern-0.75em\raise-1.1ex\hbox{$\sim$}}}

\def\beqa{\begin{eqnarray}}
\def\eeqa{\end{eqnarray}}

\begin{document}

\title{QCD Wehrl and entanglement entropies in a gluon spectator model at small--$x$}
\author{Gabriel Rabelo--Soares$^{1}$}
\email{grsoares@ifi.unicamp.br}
\author{Reinaldo Francener$^{1}$}
\email{reinaldofrancener@gmail.com}
\author{Gabriel S. Ramos $^{2}$} 
\email{gsramos.7@hotmail.com}
\author{Giorgio Torrieri$^{1}$}
\email{torrieri@unicamp.br}
\affiliation{$^{1}$ Universidade Estadual de Campinas -- Instituto de Física Gleb Wataghin\\
Rua Sérgio Buarque de Holanda, 777\\
 CEP 13083--859 -- Campinas -- São Paulo - Brazil}
\affiliation{$^{2}$ High Energy Physics Phenomenology Group, GFPAE. Institute of Physics, Federal University of Rio Grande do Sul (UFRGS)\\
Caixa Postal 15051, CEP 91501--970, Porto Alegre, RS, Brazil} 

\begin{abstract}
Recent studies have shown that hadronic multiplicity in deep inelastic scattering can be associated with entanglement entropy. However, such definitions are intrinsically longitudinal and do not capture the full phase-–space structure of the proton. In this work, we investigate the proton Wehrl entropy constructed from the gluon Husimi distribution, which provides a positive phase--space description within the
present definitions and model calculations. Within this framework, we employ a gluon light-–front spectator model based on soft–-wall AdS/QCD–-inspired wave functions, with free parameters constrained by global NNPDF fits, allowing us to compute both parton distribution functions and Wigner distributions. The Husimi distribution is obtained via Gaussian smearing of the Wigner distribution with width given by the saturation scale in the GBW model. We show that from a normalized Husimi distribution one can decompose the Wehrl entropy into an entanglement entropy term and a residual term associated with transverse degrees of freedom. Numerical results for the proton entanglement entropy are shown and compared with CMS data, while the Wehrl entropy is presented for different values of the virtuality.
\end{abstract}
\maketitle

\section{Introduction}

Currently, a significant number of studies have focused on the relevance of quantum--information concepts applied to physical observables of interest in high--energy physics. In particular, quantum entanglement has emerged as a central principle in the entropic characterization of partonic systems. In this context, the Kharzeev--Levin (KL) model~\cite{kharzeev2017deep} plays a fundamental role by establishing, through a simple relation, $S_{EE} = \ln N$, a connection between the entanglement entropy $S_{EE}$ and the effective number of partonic degrees of freedom $N$, which can be inferred from parton distribution functions (PDFs). This model successfully describes the hadronic entropy extracted from multiplicity distributions~\cite{tu2020einstein,kharzeev2021deep,hentschinski2022evidence,hentschinski2022maximally,kutak2024entanglement,hentschinski2024qcd,hentschinski2025entanglement} and provides relevant predictions for future experiments at the Electron--Ion Collider (EIC)~\cite{Ramos,Accardi:2012qut,AbdulKhalek:2021gbh}. Such agreement is particularly striking given that PDFs encode information about the initial partonic state, whereas hadronic multiplicity is a final--state observable after hadronization.

Moreover, it is worth highlighting the entropic definition introduced by Alfred Wehrl~\cite{wehrl1979relation}, which provides a semiclassical construction in phase--space based on a minimal coarse--graining compatible with the uncertainty principle. This entropic notion aims at quantifying the minimal amount of information that is unavoidably lost when a quantum system is described through effective phase--space distributions. In this context, the Wehrl entropy is defined in terms of the Husimi distribution \cite{Husimi1940}, which for the well defined entropy corresponds to a smoothed and positive--definite version of the Wigner distribution \cite{Wigner32,wignerji} and explicitly incorporates the quantum limitation on phase--space resolution.

In the small--$x$ regime of quantum chromodynamics (QCD), gluon Wigner and Husimi distributions can be formulated in the natural phase--space spanned by the Bjorken variable \(x\), the impact parameter $\mathbf{b}_{\perp}$, and the transverse momentum $\mathbf{k}_{\perp}$~\cite{hagiwara2018classical}. In this limit, the gluon Wigner distribution is approximately related to the dipole $S$--matrix, whose evolution is governed by the Balitsky--Kovchegov (BK) equation, allowing saturation effects to be incorporated in a controlled manner~\cite{PhysRevD.94.094036}. However, in the major cases, the Wigner distribution is not positive definite and exhibits sensitivity to infrared and confinement effects. The Husimi distribution, obtained via an appropriate minimal Gaussian smearing in $\mathbf{b}_{\perp}$ and $\mathbf{k}_{\perp}$ \cite{hagiwara2015use, hatta2016husimi}, circumvents these issues, yielding a positive and well--defined phase--space distribution. This construction is particularly natural in the context of the Color Glass Condensate (CGC)~\cite{gelis2010color}, whose formulation is inherently based on coherent gluon states. Such phase--space distributions remain relatively scarce in the literature and are closely connected to hadronic tomography programs~\cite{hagiwara2015use, hatta2016husimi}, potentially providing access to additional observables, such as angular correlations and the orbital angular momentum of partons.

In this context, it is useful to introduce the light--front spectator model for the proton~\cite{Chakrabarti:2023djs,Chakrabarti:2025qba}, in which the partonic structure of the proton is effectively described as a system composed of an active gluon and a spin--$\frac{1}{2}$ spectator. Within this framework, the proton is expanded in two--body Fock states, with the dynamics encoded in light--front wave functions that depend on the longitudinal momentum fraction $x$ and the transverse momentum of the gluon. The wave functions are constructed using a parametrization inspired by the soft--wall AdS/QCD formalism, incorporating free parameters associated with the longitudinal profile, the transverse width, and the overall normalization of the state. These parameters are fixed by fitting the unpolarized gluon distribution to global phenomenological parametrizations, ensuring a consistent description of the gluon PDF in the kinematic regime of interest. Once the light--front wave functions are determined, the model provides a unified framework for the construction of collinear PDFs, transverse momentum distributions (TMDs), generalized parton distributions (GPDs), and, through appropriate Fourier transform of the generalized transverse momentum distributions (GTMDs), gluon Wigner distributions, thereby enabling direct access to the phase--space structure of the proton.

The advantages offered by the light--front spectator model for the proton is that it provides an ideal framework for a comparative study of different entropic notions in QCD. In particular, the entanglement entropy within the KL model, which depends exclusively on the collinear parton distribution, can be directly obtained from the light--front spectator model once the parameters are fixed through the fit of the gluon PDF. Using exactly the same set of parameters, it is then possible to construct the corresponding Wigner distribution in phase--space and, from it, define a Husimi distribution via a minimal Gaussian smearing, allowing for the evaluation of the semiclassical Wehrl entropy. In this way, the present work aims at investigating how semiclassical phase--space entropies behave in the small--$x$ regime of QCD, where gluons dominate the dynamics and effective descriptions based on saturation become particularly relevant.

The paper is organized as follows. In Section~\ref{sec:theoretical_formalism_of_high_energy_entropy_models}, we present the theoretical formalism of high--energy entropy models, discussing the entanglement entropy in the Kharzeev--Levin framework and the formulation of Wehrl entropy in QCD based on phase--space distributions. In Section~\ref{sec:pdf_and_wigner_distributions_in_the_CGMS_phenomenology}, we describe the light--front gluon spectator model and its use to construct gluon parton distribution functions and Wigner distributions within a unified framework, with parameters fixed through fits to phenomenological PDF data. In Section~\ref{sec:results}, we present the results for the entanglement and Wehrl entropies, compare them with experimental data, and analyze their behavior in the small--$x$ regime. Finally, in Section~\ref{sec:conclusion}, we summarize our main findings and discuss the relevance of the Wehrl entropy for the entropy content of the proton.

\section{THEORETICAL FORMALISM OF HIGH ENERGY ENTROPY MODELS}
\label{sec:theoretical_formalism_of_high_energy_entropy_models}
\subsection{The Kharzeev--Levin model of the high--energy entanglement entropy}
\label{subsec:the_KL_model_of_the_high_energy_entanglement_entropy}
In the high--energy gluon physics framework, the investigation of entanglement entropy (EE) was fundamentally motivated by the Kharzeev--Levin model~\cite{kharzeev2017deep}.
Here, this model will be briefly presented in its original form, which is sufficient for what is intended to be demonstrated. To understand it, as well as its relation to Wehrl entropy, we start from classical mechanics, in which the Boltzmann-–Gibbs entropy is given by \cite{Gibbs1902}
\begin{equation}
S_{BG}=-k_B\int \frac{dpdq}{h^\prime}f(q,p)\ln f(q,p),
\label{bg-entropy}
\end{equation}
with $q$ and $p$ the generalized coordinate and momentum variables. In this equation, $h^\prime$ is an elementary cell in the classical phase--space of the system described by the function $f(q,p)$. Otherwise, in quantum mechanics, the entropy is computed from the {\it von Neumann} expression \cite{vonNeumann1955}
\begin{equation}
S_{vN}=-k_B\Tr[\hat \rho\ln \hat \rho],
\label{vn_entropy}
\end{equation}
with $\hat{\rho}$ being the density operator. Instead of a statistical distribution in phase--space, the entropy is now calculated from the density operator $\hat{\rho}$ of a given quantum system. For a bipartite entangled system, the general physical state $\ket{\psi} \in \mathcal{H}$ can be expressed, via the Schmidt decomposition, as
\begin{equation}
\ket{\psi}=\sum_{i}\alpha_i\ket{a_i}\otimes\ket{b_i}.
\label{general_entangled_state}
\end{equation}
The entangled subsystems $A$ and $B$ define their respective orthonormal bases, $\{\ket{a_i}\}\in\mathcal{H}_A$ and $\{\ket{b_i}\}\in\mathcal{H}_B$, with $\mathcal{H}_A\otimes\mathcal{H}_B=\mathcal{H}$.

The density matrix of the state of Eq. (\ref{general_entangled_state}) is given by:
\begin{equation}
\hat{\rho} = \sum_{i,j} \alpha_i \alpha_j^{*}\ket{a_i}\bra{a_j}\otimes\ket{b_i}\bra{b_j}.
\label{density_matrix_entangled}
\end{equation}
Applying the partial trace over subsystem $B$, one obtains the reduced density matrix $\hat \rho_A$:
\begin{equation}
\hat \rho_A = \Tr_B \hat \rho = \sum_i \alpha_i^2 \ket{a_i}\bra{a_i}.
\label{partial_trace}
\end{equation}
Conversely, if the partial trace were taken over subsystem $A$, one would obtain $\hat \rho_B=\sum_i \alpha_i^2\ket{b_i}\bra{b_i}$. From this, the EE is expressed as:
\begin{equation}
S_{EE}=-\Tr[\hat \rho_A\ln\hat \rho_A].
\label{ee_entropy}
\end{equation}
This expression measures the degree of entanglement of a system, and a remarkable property is $S(\hat \rho_A)=S(\hat \rho_B)$.

In general, the main challenges in evaluating EE consist in computing the logarithm in the Eq. (\ref{ee_entropy}), with various strategies existing for this purpose. For example, the EE between scattered particles in elastic collisions within the S--matrix formalism performs this computation through the Rényi entropy $S_\gamma$, using the fact that $\lim_{\gamma\rightarrow 1}S_\gamma=S_{EE}$~\cite{peschanski2016entanglement,peschanski2019evaluation}. Another way to perform this evaluation is by using the definition of the logarithm, $\ln \hat A=\lim_{\epsilon\rightarrow0}\frac{1}{\epsilon}(\hat A^\epsilon-\hat 1)$, as used in the EE in the CGC~\cite{kovner2015entanglement}. The KL model uses a simpler approach evaluating entanglement from Shannon entropy~\cite{Shannon:1948dpw}, $S_{EE}=-\sum_ip_n\ln p_n$ with $p_n\equiv \alpha_n^2$.

Thus, in the original KL model approach, the system is divided into two sets: the measured states in a deep inelastic scattering (DIS), $A$, and the unmeasured part, $B$, such that $A \cup B$ composes the entire hadron. The entanglement information of this system comes from $p_n(Y)$, where $Y$ is the rapidity. The most simplified version of the problem evaluates this dynamics in one spatial and one energetic dimension $(1+1)$ in a color dipole cascade:
\begin{equation}
\frac{d p_n}{dY} = -\omega_0 n p_n + (n-1)\,\omega_0 p_{n-1}.
\label{cascate_equation}
\end{equation}
In Eq.~(\ref{cascate_equation}), $p_n(Y)$ is the probability of finding $n$ dipoles with rapidity $Y$, and $\omega_0$ is the  Balitsky--Fadin--Kuraev--Lipatov (BFKL) intercept that computes the probability of a dipole decaying into two \cite{Mueller:1989st,Mueller:1994jq}. Moreover, the cascade dynamics emerges from the Balitsky-–Kovchegov equation \cite{Balitsky:1995ub,Kovchegov:1999yj}.

The boundary conditions for this equation revolve around the normalization of probabilities, $\sum_ip_n=1$, and the fact that at $Y=0$, there is only one dipole, $p_1(0)=1$, such that $p_{n>1}(0)=0$. Following the methodology of A. Mueller~\cite{mueller1995unitarity}, the solution of this equation is:
\begin{equation}
p_n(Y)=e^{-\omega_0 Y}(1-e^{-\omega_0 Y})^{\,n-1}.
\label{solution_cascate}
\end{equation}
With this expression, one simply substitutes it into the EE of Eq.~(\ref{ee_entropy}) and, in the high--energy limit $S\approx \omega_0 Y$. Now, it is possible to compute the mean number of particles:
\begin{equation}
xG(x)=\sum_n np_n=e^{\omega_0 Y}.
\label{gluons_distributions}
\end{equation}
Thus, the EE in high--energy physics is given by a simple expression:
\begin{equation}
S(x)=\ln [xG(x)],
\label{kl_model_entropy}
\end{equation}
with $xG(x)$ representing the gluon distribution function.
With the evolution of the state--of--the--art, it is known that besides the contribution from gluons, the inclusion of the sea--quark distribution is required~\cite{hentschinski2022maximally,hentschinski2022evidence}. However, in the present work, this fact will be ignored, as gluons alone are sufficient for the purposes intended here. It is worth noting that the original formulation of the KL model effectively captures the high--energy dynamics in terms of a one--dimensional evolution in rapidity. This framework was subsequently generalized to incorporate transverse dipole degrees of freedom and impact--parameter dependence within the QCD dipole cascade and CGC formalisms, where the typical transverse scale is set by the dipole size or, equivalently, by the saturation momentum. Remarkably, despite the inclusion of these additional degrees of freedom, the resulting entanglement entropy retains the universal form, confirming the robustness of the KL result beyond the simplified $(1+1)$--dimensional picture~\cite{gotsman2020high}.

\subsection{QCD Wehrl entropy}
\label{sec:the_qcd_wehrl_entropy}

To understand the formulation of Wehrl entropy in QCD, we evaluate the arguments proposed in reference~\cite{hagiwara2018classical}. The transition from a quantum entropic description to a classical one is not trivial. That is, the Eq.~(\ref{vn_entropy}) does not reduce to the Eq.~(\ref{bg-entropy}) in the limit $\hbar\rightarrow 0$. To investigate how this evolution takes place, A. Wehrl proposed a notion of semiclassical entropy~\cite{wehrl1979relation}. To understand it, consider the coherent state $\ket{\lambda}$, with eigenvalue $\lambda=\frac{1}{\sqrt{2h}}(q+ip)$ of the annihilation operator, $a\ket{\lambda}=\lambda\ket{\lambda}$. Taking the trace of the {\it von Neumann} entropy defined in Eq.~(\ref{vn_entropy}), one obtains:
\begin{equation}
S_{vN}=-\int \frac{dqdp}{h}\bra{\lambda}\hat{\rho}\ln \hat{\rho}\ket{\lambda}.
\label{vn_coherent_state}
\end{equation}

The Wehrl entropy, $S_W$, is obtained by performing the {\it classical substitution}, which consists of replacing $\bra{\lambda}\hat{\rho}\ln \hat{\rho}\ket{\lambda}$ with $\bra{\lambda}\hat{\rho}\ket{\lambda}\,\ln \bra{\lambda}\hat{\rho}\ket{\lambda}$. Thus,
\begin{equation}
S_W=-\int \frac{dqdp}{h}\bra{\lambda}\hat{\rho}\ket{\lambda}\,\ln \bra{\lambda}\hat{\rho}\ket{\lambda}.
\label{classical_wehrl_entropy}
\end{equation}
Since $-x\ln x$ is a concave function, using Jensen’s inequality one finds $S_W>S_{vN}\geq 0$, and the equality $S_W=S_{vN}$ is impossible; $S_W$ is always nonzero even for a pure state.

Now, considering a one--dimensional quantum system with a generic time--dependent pure state $\ket{\psi(t)}$, the Wigner distribution is defined as
\begin{equation}
W(q,p,t)
= 
\int_{-\infty}^{\infty} dx\, e^{-ipx/\hbar}\,
\bra{\, q + x/2\,}\hat{\rho}(t)\ket{\,q - x/2\,},
\label{wigner_definition}
\end{equation}
where $\hat{\rho}(t)$ is the density matrix of a pure state. The Wigner distribution is a function of both position $q$ and momentum $p$, satisfying
\begin{equation}
\begin{cases}
\displaystyle \int dq\, \frac{1}{2\pi\hbar} W(q,p,t) = \left| \bra{\psi(t)} p \rangle \right|^2 ,\\[6pt]
\displaystyle \int dp\, \frac{1}{2\pi\hbar} W(q,p,t) = \left| \bra{\psi(t)} q \rangle \right|^2, \\[6pt]
\displaystyle \int dq\, dp\, \frac{1}{2\pi\hbar} W(q,p,t) = 1,
\end{cases}
\label{wigner_properties}
\end{equation}
with the last property representing normalization. The set of properties~\eqref{wigner_properties} makes it tempting to interpret $W$ as a probability distribution in phase--space $(q,p)$. However, the Wigner distribution is highly oscillatory and not positive--definite. This occurs due to the uncertainty principle. Thus, the best one can do is describe the system by probabilities of finding the particle inside the band $(q \pm \sigma_q/2, p \pm \sigma_p/2)$, with minimal uncertainty $\sigma_q \sigma_p = \hbar/2$. This property is satisfied by the Husimi distribution, which can be obtained from the Gaussian convolution of the Wigner distribution:
\begin{equation}
H(q,p,t)=\int\frac{dq'dp'}{\pi\hbar}e^{-\frac{m\omega(q-q')^{2}}{\hbar}-\frac{(p-p')^{2}}{m\omega\hbar}}
W(q',p',t).
\label{husimi_definition}
\end{equation}
This expression is also known as the Weierstrass transform. Here, $m$ is the particle’s mass and $\omega$ is an arbitrary parameter; for oscillating systems, including radiation fields, $\omega$ is identified as the frequency.

If the gaussian smearing suppresses the oscillatory and negative features of the Wigner distribution, the Husimi distribution will be positive semi--definite:
\begin{equation}
H(q,p,t)\equiv \bra{\lambda} \hat{\rho} \ket{\lambda} = \left|\langle \psi | \lambda \rangle\right|^{2} \ge 0,
\label{husimi_positive}
\end{equation}
This expression is essentially the trace of the density matrix in the $\lambda$ basis. Using the definition of Eq.~\eqref{husimi_positive}, one may write Wehrl entropy as:
\begin{equation}
S_W = -\int \frac{dq\, dp}{h}\, H(q,p)\ln H(q,p).
\label{wehrl_entropy}
\end{equation}
Based on Eq.~\eqref{wigner_definition}, it is also possible to define an alternative entropic quantity:
\begin{equation}
\bar{S}_W = -\int \frac{dq\, dp}{h}\, W(q,p)\ln W(q,p).
\label{alternative_wehrl}
\end{equation}
However, this expression can only be evaluated in the rare cases where the Wigner distribution is positive--definite.

In the high--energy regime, partons are characterized by the Bjorken variable $x$, the transverse momentum $\mathbf{k}_\perp$, and the impact parameter $\mathbf{b}_\perp$. Thus, the characterization of the Wigner distribution is built upon this set of variables, and in QCD the relation with the Husimi distribution is given by:
\begin{equation}
\begin{aligned}
    &xH(x,\mathbf{b}_{\perp},\mathbf{k}_{\perp}) =\frac{1}{\pi^2} \int d^{2}\mathbf{b}_{\perp}'d^{2}\mathbf{k}_{\perp}'  e^{-(\mathbf{b}_{\perp}-\mathbf{b}_{\perp}')^{2}/\ell^{2} - (\mathbf{k}_{\perp}-\mathbf{k}_{\perp}')^{2}\ell^{2}}xW(x,\mathbf{b}_{\perp}',\mathbf{k}_{\perp}').
\end{aligned}
\label{qcd_weistrass_transform}
\end{equation}
In this equation, $\ell$ is a free parameter and, in this work, it is chosen to be the inverse saturation scale, $\ell = 1/Q_s(x)$~\cite{hagiwara2015use, hatta2016husimi}. This choice reflects the fact that the phase--space coarse--graining scale should be set by the typical transverse momentum of gluons in the hadron. While for $x \sim \mathcal{O}(1)$ this scale is nonperturbative and of order $\Lambda_{\mathrm{QCD}}$, in the high--energy small--$x$ regime a new perturbative scale emerges, namely the saturation momentum $Q_s(x)$, which governs the transition between dilute and dense gluonic systems. In this regime, the CGC provides the appropriate effective description, and the identification $1/\ell = Q_s(x)$ becomes natural, since $\langle k_\perp \rangle \sim Q_s(x)$ sets the typical transverse momentum scale of the system. A phenomenological determination of the saturation scale is provided by the Golec--Biernat--W\"usthoff (GBW) model~\cite{GBWI,GBWII}, which parametrizes saturation effects through $Q_s^2(x) = Q_0^{2}(x_0/x)^{\lambda}$, with the parameters $x_0$, $\lambda$ and $Q_0^{2}$ extracted from deep inelastic scattering data in $ep$ collisions. In this work, we adopt representative values $x_0 = 4.2 \times 10^{-5}$, $\lambda = 0.248$ and $Q_0^{2} =  1 \, \text{GeV}^{2}$, which are known to provide a successful description of small--$x$ DIS data~\cite{golec2018saturation}.

Thus, Wehrl entropy in QCD is given by:
\begin{equation}
S_{W}(x) \equiv - \int d^{2}\mathbf{b}_{\perp}\, d^{2}\mathbf{k}_{\perp}\, xH(x,\mathbf{b}_{\perp},\mathbf{k}_{\perp})\, \ln[xH(x,\mathbf{b}_{\perp},\mathbf{k}_{\perp})],
\label{wehrl_qcd}
\end{equation}
and one also keeps the definition of $\bar{S}_{W}(x)$ using the Wigner distribution,
\begin{equation}
\bar{S}_{W}(x) \equiv - \int d^{2}\mathbf{b}_{\perp}\, d^{2}\mathbf{k}_{\perp}\, xW(x,\mathbf{b}_{\perp},\mathbf{k}_{\perp})\, \ln[xW(x,\mathbf{b}_{\perp},\mathbf{k}_{\perp})].
\label{wehrl_wigner}
\end{equation}
Thus, the characterization of Wehrl entropy in QCD depends on establishing a Husimi distribution. In the next sections, the formalism that will provide the Wigner distribution and hence the Husimi distribution will be developed.

\section{PDF and Wigner distributions in the gluon spectator model}
\label{sec:pdf_and_wigner_distributions_in_the_CGMS_phenomenology}

In this work, we employ the light--front wave function (LFWF) formalism within the gluon light--front spectator model~\cite{Chakrabarti:2023djs}, suitably adapted so that the effective two--particle wave function is obtained from the soft--wall AdS/QCD framework~\cite{Brodsky:2014yha,Chakrabarti:2025qba}. This approach provides a unified description of the longitudinal and transverse structure of the proton, allowing not only for the construction of collinear parton distribution functions, but also for the definition of more general phase--space distributions. In particular, within this framework one can access gluon Wigner distributions, originally introduced in quantum mechanics by Wigner~\cite{Wigner32} and later extended to QCD by Ji~\cite{wignerji}, which encode simultaneously the longitudinal momentum fraction $x$ and the transverse degrees of freedom, namely the parton transverse momentum $\mathbf{k}_\perp$ and the impact parameter $\mathbf{b}_\perp$. These Wigner distributions act as generating functions for TMDs and GPDs, which are recovered as marginal distributions. Although Wigner distributions are not directly measurable, they are related to GTMDs, from which they can be obtained via a Fourier transform in the transverse momentum transfer in the skewness limit as will later be done in this work. In the present work, we do not attempt to rederive the formalism for GTMDs or Wigner distributions developed in Refs.~\cite{Chakrabarti:2023djs,Chakrabarti:2025qba,Sain:2025kup}. Instead, we focus on employing this established framework to consistently construct both PDFs and phase--space distributions, which will later be used to investigate the entropy content of the proton.

The light--front gluon spectator model is a way of modeling the proton incorporating the gluons as a degree of freedom. Therefore, in the high energy scattering of the proton regime, the gluon is treated as the active parton as the rest of the proton treated as spin--$\frac{1}{2}$ fermion spectator with a effective mass. The proton is constructed from a light--cone wave function using the AdS/QCD prediction with the free parameters obtained from the fit of the model with the NNPDF4.0nnlo data set \cite{NNPDF:2021njg}.

At leading order, the Fock state of the proton is composed by the valence quarks only. Including higher order terms on the Fock decomposition, one have the gluons and sea quarks. The first higher order Fock state for the proton that contains a gluons is $| qqq g \rangle$. This is a state of a four body system, which is very complicated to work with. The spectator model describe the proton as a state formed by on active gluon and a spin--$\frac{1}{2}$ spectator \cite{Lu:2016vqu}, i.e., the proton is a composed state of the form, 
\begin{equation}
    |P;S \rangle \mapsto | g^{s_g} X^{s_X}(uud) \rangle,
    \label{fockspectator}
\end{equation}
with $s_g = 1$ representing the spin of the gluon, and $s_{X} = \frac{1}{2}$ representing the spin of the proton (the spectator particle). In the proton frame, its momentum is $P = (P^{+}, M^{2}/P^{+}, \mathbf{0}_\perp)$. The momentum of the gluon (active parton) $p= (xP^{+}, \frac{p^{2} + \mathbf{k}_\perp^{2}}{xP^{+}}, \mathbf{k}_\perp)$ and the spectator momentum $P_X = ((1-x)P^{+}, P^{-}_X, - \mathbf{k}_\perp)$ with the longitudinal momentum carried by the parton being $x = p^{+}/P^{+}$. Therefore, the proton state as in Eq. (\ref{fockspectator}) can be written as a two--particle Fock--State expansion with proton spin components $J_z = \pm \frac{1}{2}$ \cite{Brodsky:2000ii}, 
\begin{equation}
    \begin{aligned}
        \left| P; \uparrow (\downarrow) \right\rangle =
\int \frac{d^2 \mathbf{k}_\perp \, dx}{16\pi^3\, x(1-x)}
& \Big[
\psi^{\uparrow(\downarrow)}_{+1\,+\frac{1}{2}}(x,\mathbf{k}_\perp)\,
\big| +1, +\tfrac{1}{2};\, xP^+, \mathbf{k}_\perp \big\rangle
 + \psi^{\uparrow(\downarrow)}_{+1\,-\frac{1}{2}}(x,\mathbf{k}_\perp)\,
\big| +1, -\tfrac{1}{2};\, xP^+, \mathbf{k}_\perp \big\rangle \\
&
\quad \psi^{\uparrow(\downarrow)}_{-1\,+\frac{1}{2}}(x,\mathbf{k}_\perp)\,
\big| -1, +\tfrac{1}{2};\, xP^+, \mathbf{k}_\perp \big\rangle
\psi^{\uparrow(\downarrow)}_{-1\,-\frac{1}{2}}(x,\mathbf{k}_\perp)\,
\big| -1, -\tfrac{1}{2};\, xP^+, \mathbf{k}_\perp \big\rangle
\Big],
    \end{aligned}
    \label{protondec}
\end{equation}
where $\psi^{\uparrow(\downarrow)}_{\lambda_g \lambda_X}(x,\mathbf{k}_\perp)$ are the light--front wave functions for the two particle state $\big|\lambda_g, \lambda_X;\, xP^+, \mathbf{k}_\perp \big\rangle $ with proton helicities $\lambda_p = \uparrow(\downarrow)$. Following the model developed by the authors in \cite{Chakrabarti:2023djs}, the LFWF in Eq. (\ref{protondec}) are inspired by the wave function of the physical electron as \cite{Brodsky:2000ii}, and so, the light--front wave functions for the Fock--state expansion (Eq. (\ref{protondec})) for a proton with $J_z = + 1/2$ is,
\begin{equation}
\begin{aligned}
\psi^{\uparrow}_{+\frac{1}{2}\,+1}(x,\mathbf{k}_\perp) 
&= -\sqrt{2}\,\frac{-\mathbf{k}_{\perp}^{(1)}+i \mathbf{k}_{\perp}^{(2)}}{x(1-x)}\,
\varphi(x,\mathbf{k}_\perp^{2}), \\[4pt]
\psi^{\uparrow}_{+\frac{1}{2}\,-1}(x,\mathbf{k}_\perp) 
&= -\sqrt{2}\,
\left(\frac{M-M_X}{1-x}\right)\,
\varphi(x,\mathbf{k}_\perp^{2}), \\[4pt]
\psi^{\uparrow}_{-\frac{1}{2}\,+1}(x,\mathbf{k}_\perp) 
&= -\sqrt{2}\,
\frac{\mathbf{k}_{\perp}^{(1)}+i \mathbf{k}_{\perp}^{(2)}}{x}\,
\varphi(x,\mathbf{k}_\perp^{2}), \\[4pt]
\psi^{\uparrow}_{-\frac{1}{2}\,-1}(x,\mathbf{k}_\perp) 
&= 0 ,
\end{aligned}
\end{equation}
and for $J_z = - 1/2$,
\begin{equation}
\begin{aligned}
\psi^{\downarrow}_{+\frac{1}{2}\,+1}(x,\mathbf{k}_\perp) 
&= 0, \\[4pt]
\psi^{\downarrow}_{+\frac{1}{2}\,-1}(x,\mathbf{k}_\perp) 
&= -\sqrt{2}\,
\frac{-\mathbf{k}_{\perp}^{(1)}+ i \mathbf{k}_{\perp}^{(2)}}{x}\,
\varphi(x,\mathbf{k}_\perp^{2}), \\[4pt]
\psi^{\downarrow}_{-\frac{1}{2}\,+1}(x,\mathbf{k}_\perp) 
&= -\sqrt{2}\,
\left(\frac{M-M_X}{1-x}\right)
\varphi(x,\mathbf{k}_\perp^{2}), \\[4pt]
\psi^{\downarrow}_{-\frac{1}{2}\,-1}(x,\mathbf{k}_\perp) 
&= -\sqrt{2}\,
\frac{\mathbf{k}_{\perp}^{(1)}+ i \mathbf{k}_{\perp}^{(2)}}{x(1-x)}\,
\varphi(x,\mathbf{k}_\perp^{2}) .
\end{aligned}
\end{equation}
In this construction, $\varphi(x,\mathbf{k}_\perp)$ is inspired by the soft--wall AdS/QCD wave function \cite{Gutsche:2013zia} with two additional parameters, \textit{a} and \textit{b}, in order to regulate the asymptotic behavior of the gluon PDF. The complete form of the wave function is given by
\begin{equation}
    \varphi\left(x, \mathbf{k}_\perp^{\,2}\right)
=
\frac{N_g}{4\pi\,\kappa}
\sqrt{\frac{\log\!\left(\tfrac{1}{1-x}\right)}{x}}
\, x^{\,b}(1-x)^{\,a}
\exp\!\left[
-\frac{\log\!\left(\tfrac{1}{1-x}\right)}{2\kappa^{2}x^{2}}
\,\mathbf{k}_\perp^{\,2}
\right]\, .
\label{softwall}
\end{equation}
In this formalism, $N_g$, $a$ and $b$ are free parameters fixed by fitting the unpolarized gluon PDF at some virtuality (in this work, the analysis is set for the virtualities $Q = 2,5,10$ GeV), with NNPDF4.0 at next--to--next--leading order data \cite{NNPDF:2021njg}. In this formalism $\kappa$ is an emergent mass scale parameter that describes the transverse dynamics of gluons within a hadron \cite{Sain:2025kup}.

\subsection{Model parameters and numerical fitting}
\label{model_params_num_fit}
\begin{figure}[t!]
    \centering
    \includegraphics[scale=0.7]{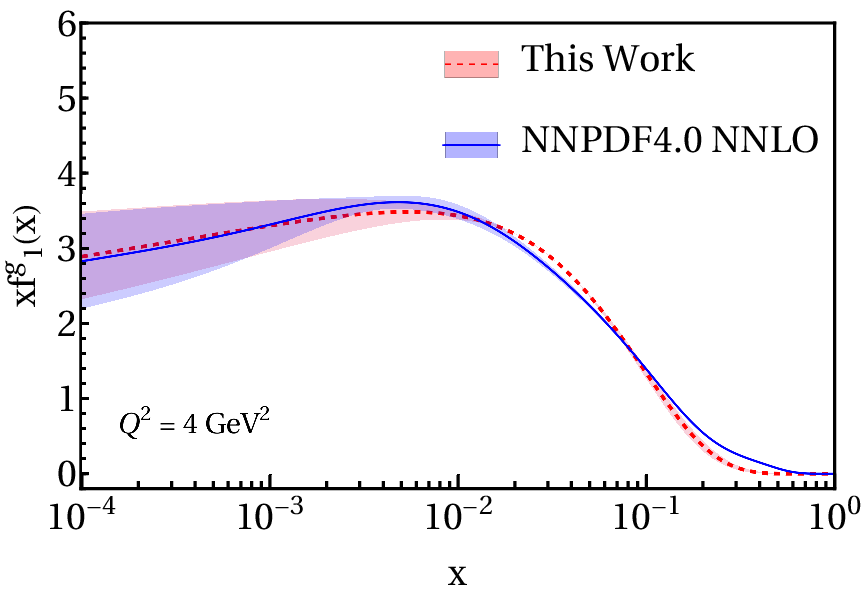}
    \caption{Unpolarized gluon PDF $f^{1}_g(x)$ as a function of the longitudinal momentum fraction $x$, in the range $0.0001 < x < 1$ for $Q^{2} = 4$ GeV$^{2}$. The solid blue line and the blue shadow represents the NNPDF 4.0 at NNLO data set for the unpolarized gluon PDF and its uncertainty band with 1$\sigma$, respectively. The dashed red line and the red shadow represents the fit of the Eq. (\ref{softwall}) with the free parameters $N_g$, $a$, and $b$ and the uncertainty band respectively.}
    \label{fitq2}
\end{figure}

As previously discussed, the wave function in the AdS/QCD inspired model is constructed in such a way that its asymptotic form recovers the standard unpolarized gluon parton distribution function. As can be seen in Eq. (\ref{softwall}), $a$ and $b$ are included in order do capture the small--$x$ feature where the PDF follow a power--law behavior $x^{b}$, motivated by Regge theory and the high--energy scattering data, while at large $x$ they are constrained by QCD power counting rules $(1-x)^{a}$ \cite{Brodsky_1995}. Therefore, the unintegrated gluon correlation function for the leading twist gluon transverse momentum distribution in a semi--inclusive process was done by the authors in \cite{Mulders:2000sh, Chakrabarti:2023djs} with the unpolarized gluon TMD defined as in Equations (9), (10) and (11) in the reference \cite{Chakrabarti:2023djs}
\begin{equation}
f_1^{g}\!\left(x,\mathbf{k}_\perp^{\,2}\right)
=
\frac{N_g^{2}}{2\pi\,\kappa^{2}}
\frac{\log\!\left(\tfrac{1}{1-x}\right)}{x}\,
x^{2b}(1-x)^{2a}
 \Big[
A(x)+\mathbf{k}_\perp^{\,2} B(x)
\Big]
\exp\!\left[-C(x)\,\mathbf{k}_\perp^{\,2}\right],
\label{gluon_unpol_TMD}
\end{equation}
with
\begin{equation}
    A(x) = \left(M - \frac{M_X}{1-x}\right)^{2}, \quad B(x) = 1 + \frac{(1-x)^{2}}{x^{2}}, \quad C(x) = \frac{\log\!\left(\tfrac{1}{1-x}\right)}{\kappa^{2}x^{2}},
\end{equation}
where $M$ is the proton mass and $M_X$ is the spectator mass, in order to guarantee the stability of the proton, the hierarchy $M_X > M$ is taken. Finally, from Eq. (\ref{gluon_unpol_TMD}), the collinear unpolarized PDF, $f_1^{g}(x)$, is obtained by integrating the transverse momentum, 
\begin{equation}
\begin{aligned}
f_1^{g}(x)
&=
\int d^{2}\mathbf{k}_\perp\,
f_1^{g}\!\left(x,\mathbf{k}_\perp^{\,2}\right) \\
&=
2N_g^{2}\,
x^{2b+1}(1-x)^{2a-2}
\Bigg[
\frac{\kappa^{2}\big(1+(1-x)^{2}\big)}
{\log\!\left(\tfrac{1}{1-x}\right)}
+
\big(M(1-x)-M_X x\big)^{2}
\Bigg].
\end{aligned}
\end{equation}
With the above analytical expression for the unpolarised gluon PDF, the free parameters will then be fixed by fitting the expression with gluon PDF data at NNLO from the global analysis by the NNPDF Collaboration \cite{NNPDF:2021njg}. In this work, we are going to work with three particular values of virtuality, $Q = 2,5,10$ GeV in with the range $0.0001 < x < 1$. In particular, Fig. \ref{fitq2} shows the NNPDF4.0 at NNLO for the unpolarized gluon PDF in the solid blue line with the blue shadow representing the band uncertainty with 1$\sigma$. While in dashed red line is showed the best fit for the free parameters $N_g$, $a$ and $b$ with the shadow red band representing the uncertainty band (with 1$\sigma$) of the fit constructed.  

The parameters with the respective uncertainty are presented in the Table \ref{params_table}. The spectator mass was chosen as $M_X = 0.985$ GeV \cite{Chakrabarti:2023djs}, and the proton mass $M = 0.938$ GeV. In order to have more physical constraints, the parameter $\kappa = 2.62$ GeV as set by the authors in \cite{Sain:2025kup}, getting a better alignment with the gluon gravitational form factors (GFFs) maintaining the consistent gluon PDF (this parameter differs from Refs. \cite{Chakrabarti:2023djs,Chakrabarti:2024hwx} where $\kappa = 0.4$ GeV is determined from quark dynamics to fit the proton electromagnetic form factor \cite{Chakrabarti:2013gra}). It is also important the clarify that the gluon mass is $M_g = 0$ in the employed model.
\begin{table}[b!]
\centering
\begin{tabular}{| c | c | c |}
\hline 
Parameter & Central value & $1\sigma$ uncertainty band \\
\hline 
$a$   & $5.99$  & $\pm 0.80$ \\ 
$b$   & $-0.47$ & $\pm 0.02$ \\ 
$N_g$ & $0.43 $ & $\pm 0.05$ \\
\hline 
\end{tabular}

\caption{Best--fit values of the model parameters with their corresponding $1\sigma$ uncertainty bands. This parameters are regard for $Q^{2} = 4$ GeV$^{2}$.}
\label{params_table}
\end{table}

\subsection{Unpolarized Wigner distribution in the gluon spectator model}
The spectator model formalism and the entire discussion above can also be used to construct the Wigner distributions of gluons. This distribution is obtained by the two--dimensional Fourier transform of the gluon--gluon GTMD correlator \cite{Lorce:2013pza} at zero skewness as \cite{Meissner:2009ww,Lorce:2011ni}
\begin{align}
W^{g}_{\lambda''\lambda'}(x,\mathbf{k}_\perp,\mathbf{b}_\perp)
=
\int \frac{d^{2}\boldsymbol{\Delta}_\perp}{(2\pi)^{2}}\,
e^{-i \boldsymbol{\Delta}_\perp \cdot \mathbf{b}_\perp}
\int \frac{dz^-\, d^{2}\mathbf{z}_\perp}{(2\pi)^{3}\, x P^{+}}\,
e^{i k \cdot z}
\left.
\big\langle
P'', \lambda'' \big|
\Gamma^{ij}
F^{+i}_{a}\!\left(-\tfrac{z}{2}\right)\,
\mathcal{W}_{ab}\!\left(-\tfrac{z}{2}, \tfrac{z}{2}\right)\,
F^{+j}_{b}\!\left(\tfrac{z}{2}\right)
\big|
P', \lambda'
\big\rangle
\right|_{z^{+}=0},
\label{wigner_full_correlator}
\end{align}
where $\lvert P', \lambda' \rangle$ and $\lvert P'', \lambda'' \rangle$
denote the initial and final nucleon states with momenta $P'$ and $P''$,
and helicities $\lambda'$ and $\lambda''$, respectively. The kinematic variable definitions follows \cite{Chakrabarti:2025qba}. The tensor $\Gamma^{ij}$ denotes the leading--twist generic gluon operator, namely $\Gamma^{ij} = \delta^{ij}_{\perp} -i \epsilon^{ij}_{\perp}$, corresponding to unpolarized and longitudinally polarized gluons, respectively. The quantity $F^{\mu\nu}_{a(b)}$ represents the gluon field--strength tensor,
while $\mathcal{W}_{ab}$ denotes the Wilson line (i.e., gauge link) connecting
the two gluon field operators, ensuring the gauge invariance of the gluon correlator. Adopting the light--cone gauge, $A^{+}=0$, the Wilson line reduces to unity. The impact parameter $\mathbf{b}_{\perp}$ represents the average position with respect to the center of momentum of the proton and is the conjugated variable of the transverse momentum transfer $\mathbf{\Delta}_\perp$ \cite{Burkardt:2015qoa}. The Wigner distribution, as will be employed in this paper, can be defined for a unpolarized gluon in a unpolarized proton from Eq. (\ref{wigner_full_correlator}), (for a detailed derivation of the Wigner distribution we refer to the Section III of \cite{Chakrabarti:2025qba}), 
\begin{align}
W_{UU}(x,\mathbf{k}_\perp,\mathbf{b}_\perp) &= \frac{1}{2} \Big[ W^{g}_{\uparrow\uparrow}(x,\mathbf{k}_\perp,\mathbf{b}_\perp)
+ W^{g}_{\downarrow\downarrow}(x,\mathbf{k}_\perp,\mathbf{b}_\perp)
\Big] \nonumber\\[6pt]
&= \int \frac{d^{2}\boldsymbol{\Delta}_\perp}{(2\pi)^{2}}\,
e^{-i \boldsymbol{\Delta}_\perp \cdot \mathbf{b}_\perp}\,
\frac{2N_g^{2}}{\pi \kappa^{2}}
\Bigg[ F_1(x) \left( \mathbf{k}_\perp^{2} - \frac{(1-x)^{2}}{4}\, \boldsymbol{\Delta}_\perp^{2} \right) + F_2(x) \Bigg] \exp\!\left[-2 F_3(x)
\left( \mathbf{k}_\perp^{2} + \frac{(1-x)^{2}}{4}\, \boldsymbol{\Delta}_\perp^{2}
\right)\right],
\label{wignernpolunpol}
\end{align}
with the parametrization functions $F_a(x)$, $F_b(x)$ and $\alpha(x)$ defined as

\begin{equation}
F_1(x)
=
\left(
\frac{1+(1-x)^2}{x^2(1-x)^2}
\right)
\sqrt{\frac{
\ln\!\left(\frac{1}{1-x}\right)^{2}
}{x^2}}
x^{2b}(1-x)^{2a} \, ,
\label{fa}
\end{equation}

\begin{equation}
F_2(x)
=
\left(
M - \frac{M_X}{1 - x}
\right)^{2}
\sqrt{\frac{
\ln\!\left(\frac{1}{1-x}\right)^{2}
}{x^2}}
x^{2b}(1-x)^{2a} \, ,
\label{fb}
\end{equation}

\begin{equation}
F_3(x)
=
\frac{1}{2 \kappa^{2}x^{2}}
\ln\!\left(\frac{1}{1-x}\right) \, .
\label{alpha}
\end{equation}
Finally, we highlight that the present analysis is performed within a light--front spectator model and therefore does not incorporate evolution equations such as DGLAP, BK, or JIMWLK. The dependence on $x$ or $Q^2$ comes from the structure of the model rather than genuine perturbative evolution. In addition, this model captures the phase--space structure of an effective partonic degree of freedom, but does not include multi--parton correlations or dynamical gluon cascades. Investigating how the entropy 
observables behave once within the QCD evolution would be an interesting direction for future work.

\section{Results and discussions}
\label{sec:results}

\begin{figure}[t!]
    \centering
    \includegraphics[scale=0.4]{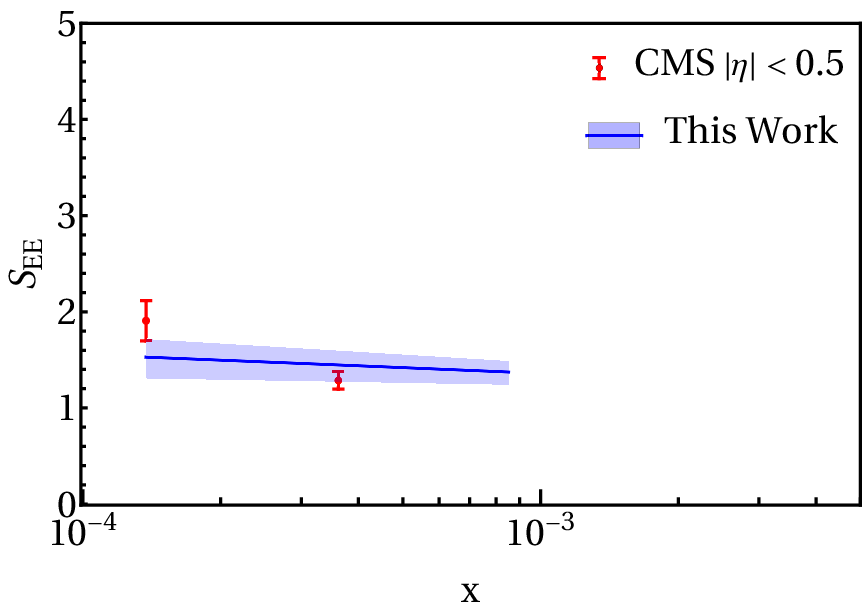}
    \includegraphics[scale=0.4]{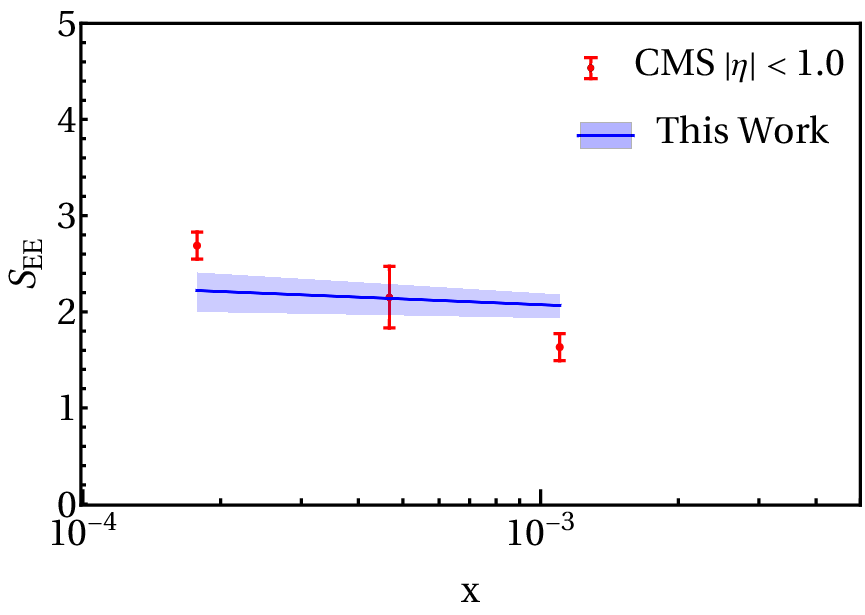}
    \includegraphics[scale=0.4]{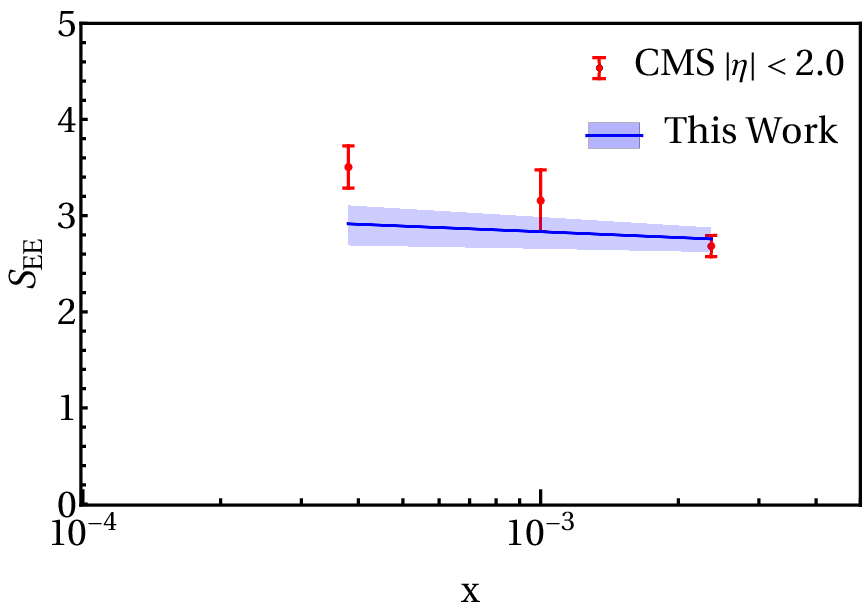}
    \caption{ The panels displays the Kharzeev--Levin entanglement entropy as a function of Bjorken--$x$ for the CMS \textit{pp} data \cite{CMS:2010qvf}, and the theoretical prediction of the model studied in this work in blue solid line. The left panel shows the entropy from the CMS data in red dot for $\sqrt{s} = 7$ TeV and  $\sqrt{s} = 2.36$ TeV with pseudo--rapidity $|\eta| < 0.5$, in blue solid line is shown this work model. The middle and right panel show the entropy from the CMS data in red dot for $\sqrt{s} = 7$ TeV, $\sqrt{s} = 2.36$ TeV and $\sqrt{s} = 0.9$ TeV with pseudo--rapidity $|\eta| < 1.0$ and $|\eta| < 2.0$ respectively, with the blue solid line the theoretical prediction.}
    \label{klmodelexp}
\end{figure}

The model studied in this work, developed by the authors in Refs. \cite{Chakrabarti:2023djs,Chakrabarti:2025qba,Sain:2025kup}, is particularly valuable because it provides, within a unified framework, both the Wigner distributions and the corresponding parton distribution functions. Since our goal is to investigate the entropy content of the proton, this model enables for the calculation of the standard entanglement entropy associated with the PDFs, which can be used as a consistency check of the framework with the experimental data \cite{tu2020einstein,Ramos}. Fig.~\ref{klmodelexp} displays the entanglement entropy extracted from the experimental CMS data \cite{CMS:2010qvf} for different pseudorapidity intervals, $|\eta|< 0.5,1.0,2.0$ and center of mass energies $\sqrt{s} = 7.00,2.36,0.90$ TeV. The blue solid lines represent the theoretical calculation of the entanglement entropy obtained using the PDF derived from Eq. (\ref{softwall}), with free parameters fitted to the NNPDF data set. To construct the plots in Fig. \ref{klmodelexp}, a similar methodology of \cite{tu2020einstein} was employed, but adapting to the saturation scale in GBW formalism.  The results show good agreement between the model predictions and the experimental data, supporting the consistency and reliability of the model. Based on this preliminary validation, the subsequent computation of the Wehrl entropy of the phase--space distribution becomes more robust.

\begin{figure}[t!]
    \centering
    \includegraphics[scale=0.5]{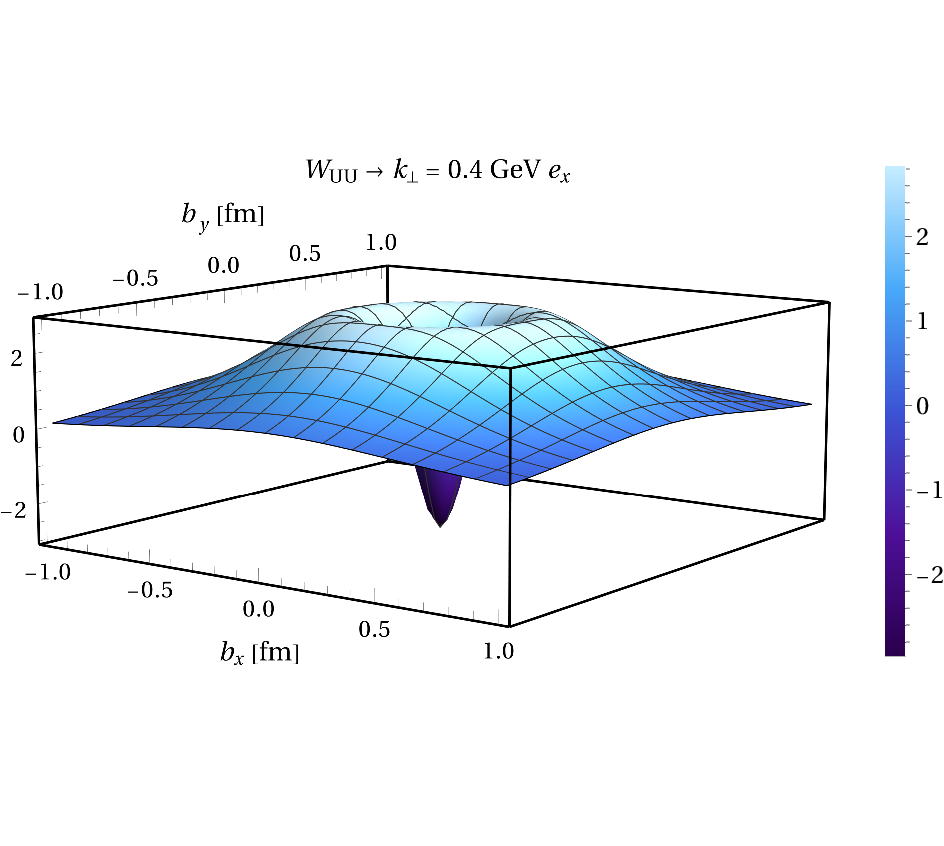}
    \includegraphics[scale=0.5]{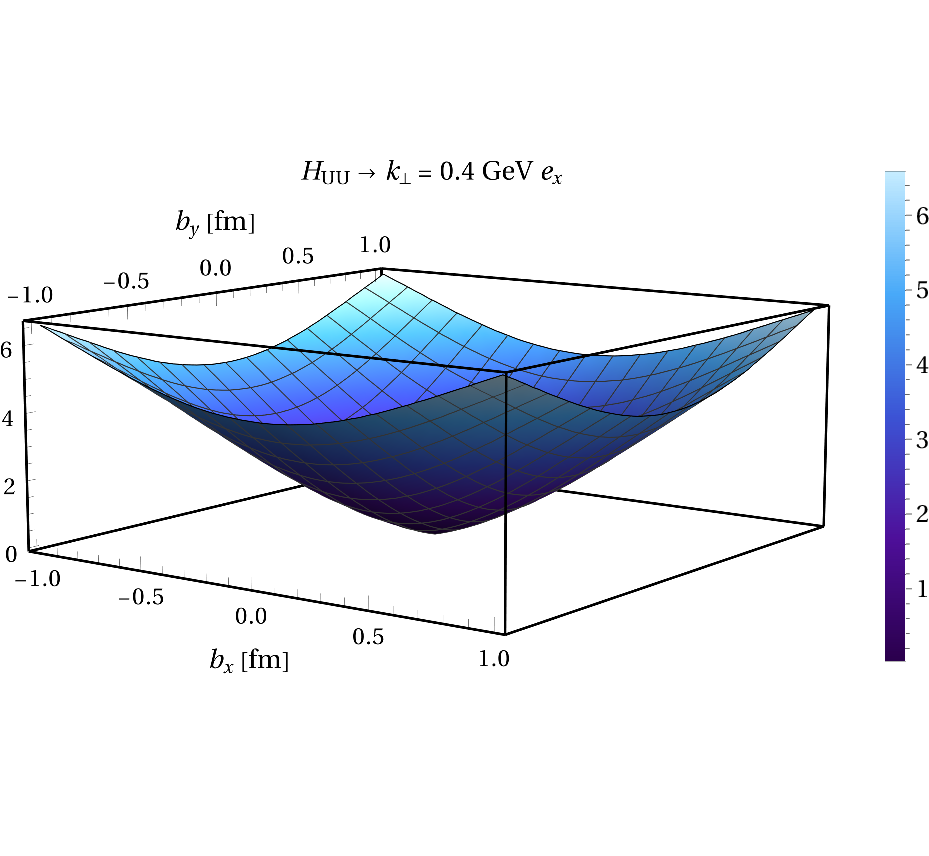}
    \caption{This figure shows the plots for the first moment of the Wigner (left) and Husimi (right) distributions for the unpolarized gluon in a unpolarized proton for a fixed transverse momentum, $k_x = 0.4$ GeV. This plots are obtained using the fit parameters from the Table \ref{params_table} for $Q^{2} = 4$ GeV$^{2}$. }
    \label{momentswh}
\end{figure}
Several physical information can be extracted from the Wigner distribution, like TMDs, GPDs, PDFs as well as spin information, OAM (orbital angular momentum), by computing the moments of the distribution. From the Wigner distribution defined in Eq. (\ref{wignernpolunpol}), in Fig. \ref{momentswh} (left panel) we are showing the first moment of the Wigner distribution from 
\begin{equation}
    W_{UU}(\mathbf{k}_\perp,\mathbf{b}_\perp) = \int dx W_{UU}(x,\mathbf{k}_\perp,\mathbf{b}_\perp).
\end{equation}
The first moment gives the effective number of parton or the density in the transverse plane. In the particular scenario of this paper, $W_{UU}(\mathbf{k}_\perp,\mathbf{b}_\perp)$ is showed in the impact parameter space for a fixed transverse momentum $\mathbf{k}_\perp = 0.4$ GeV $e_x$. As discussed in the previous sections, the positivity of the Wigner distribution is not guaranteed in the full phase--space. The consequence of non--positivity is the impossibility to define the entropy as in Eq. (\ref{wehrl_wigner}). From the Weierstrass transform defined in Eq. (\ref{qcd_weistrass_transform}), we can obtain the Husimi distribution with a resolution  $\sigma_{\mathbf{k}_{\perp}} \sigma_{\mathbf{b}_{\perp}} = 1/2$ (natural units), that follow from uncertainty principle. In order the ensure this condition, the gaussian smearing in Eq. (\ref{qcd_weistrass_transform}) with width of the gaussian set to be the saturation scale in the GBW approach \cite{GBWI, GBWII} resulted in a positive Husimi distribution throughout the phase--space within numerical accuracy. In order to make a precision comparison, the moment of the Husimi distribution for a fixed transverse momentum $\mathbf{k}_\perp = 0.4$ GeV $e_x$ is shown in Fig. \ref{momentswh} (right panel). \\

The gaussian convolution is a normalized transformation, so the Wigner distribution $W(x, \mathbf{k}_\perp, \mathbf{b}_\perp)$ and the Husimi distribution $H(x, \mathbf{k}_\perp, \mathbf{b}_\perp)$ has the same normalization and indeed both recover the parton distribution after integrating over $\mathbf{k}_\perp$ and $\mathbf{b}_\perp$,  
\begin{equation}
    \int d^{2}\mathbf{b}_{\perp}d^{2}\mathbf{k}_{\perp} xH(x,\mathbf{b}_{\perp},\mathbf{k}_{\perp}) = x f(x),
    \label{husimi_constraint}
\end{equation}
where the $xf(x)$ is the usual parton distribution function. They are not normalized to the unity but to the number of partons, in close analogy with the fact that Wigner and Husimi distributions encode phase--space densities rather than probabilities. As discussed and defined in the previous sections, the entanglement entropy in the Kharzeev--Levin model is derived from the cascade dynamics that emerges from the Balitsky--Kovchegov equation. The boundary condition relies on the normalization of the probability, i.e., $\sum^{\infty}_{n=1} p_n = 1$, which is a fundamental requirement for the definition of the Shannon entropy. Therefore, to establish a meaningful comparison between the entanglement entropy and the Wehrl entropy, the phase--space distributions must be consistently renormalized so as to satisfy an analogous probabilistic normalization.  The normalization will preserve the shape of the transverse phase--space distribution, while removes the contribution associated with the overall gluon multiplicity. Then the normalized Husimi distribution can be treated as probability density over the transverse phase--space variables \cite{hagiwara2015use, hatta2016husimi}.

\begin{figure}[t!]
    \centering
   \includegraphics[scale=0.6]{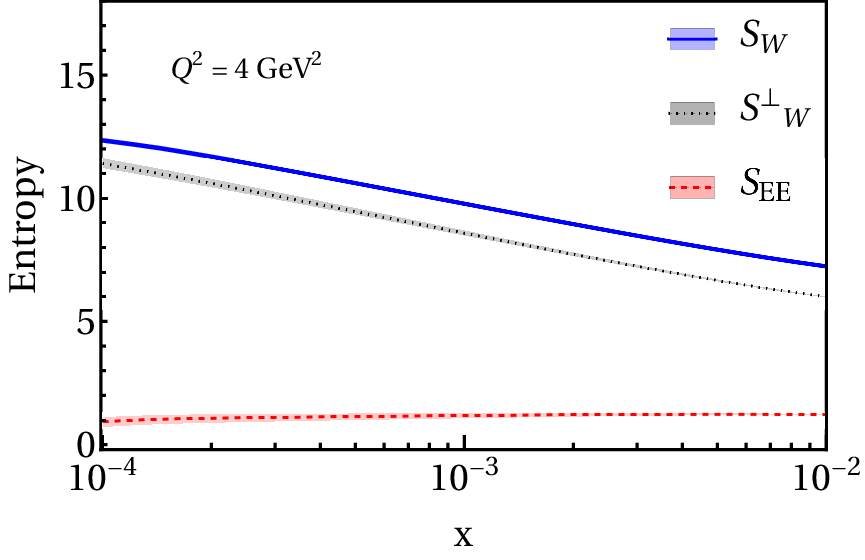}
    \includegraphics[scale=0.6]{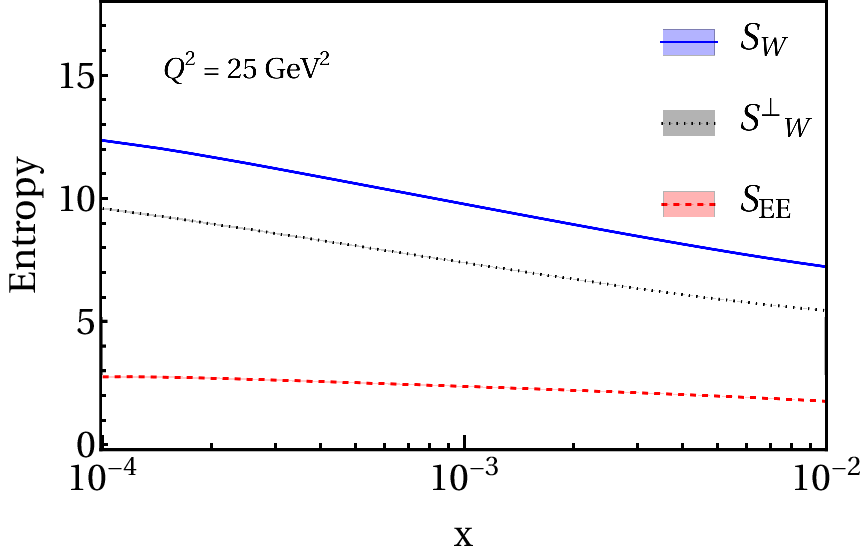}
    \includegraphics[scale=0.6]{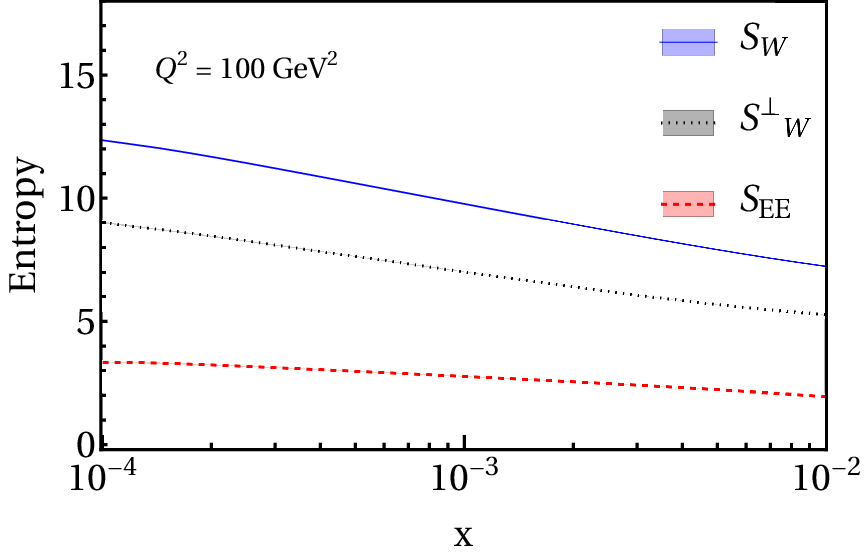}
    \caption{Plots of the Wehrl entropy as a function of the longitudinal momentum fraction as defined in Eq. (\ref{wehrldecomposition}). The full Wehrl entropy is shown by the solid blue line, the transverse entropy is shown by the dotted black line, the entanglement entropy is shown by the dashed red line. The curves are presented for three different virtuality scales: $Q = 2$  GeV (top--left panel), $Q = 5$ GeV (top--right panel), and $Q = 10$ GeV (bottom panel). The shaded bands represent the 1$\sigma$ uncertainties associated with the PDF parametrization.}
    \label{werhl_plot}
\end{figure}

Notice that the Husimi distribution no longer recovers the parton distribution as in Eq.~(\ref{husimi_constraint}), and is now written as
\begin{equation}
N(x) = \int d^{2}\mathbf{b}_{\perp}d^{2}\mathbf{k}_{\perp} xH(x,\mathbf{b}_{\perp},\mathbf{k}_{\perp}), \quad\therefore\quad  \frac{1}{N(x)}\int d^2\mathbf{k}_\perp d^2\mathbf{b}_\perp W(x,\mathbf{k}_\perp,\mathbf{b}_\perp)=1,
\label{normalization}
\end{equation}
so, the Wehrl entropy,
\begin{equation}
S_{W}(x) \equiv - \int d^{2}\mathbf{b}_{\perp}\, d^{2}\mathbf{k}_{\perp}\, \frac{xH(x,\mathbf{b}_{\perp},\mathbf{k}_{\perp})}{N(x)}\, \ln \left[\frac{xH(x,\mathbf{b}_{\perp},\mathbf{k}_{\perp})}{N(x)}\right].
\label{wherlentropy2}
\end{equation}
Some algebra can be done, noticing that the normalization $N(x)$ is indeed the parton distribution function $xf(x)$,
\begin{equation}
    \begin{aligned}
        S_{W}(x)     &\equiv -   \frac{1}{xf(x)} \int d^{2}\mathbf{b}_{\perp}\, d^{2}\mathbf{k}_{\perp}\, xH(x,\mathbf{b}_{\perp},\mathbf{k}_{\perp})  \, \ln[xH(x,\mathbf{b}_{\perp},\mathbf{k}_{\perp})] +  S_{\text{EE}}(x) \, ,
    \end{aligned}
    \label{wehrldecomposition}
\end{equation}
 one can identify the Kharzeev--Levin entanglement entropy term and the remaining term defined as $S^{\perp}_W$ (transverse Wehrl entropy) from the decomposition of the Wehrl entropy considering the normalized Husimi distribution. The term identified as $S_{\text{EE}}(x)$ is strongly dependent of the normalization condition set in Eq.~(\ref{normalization}), but argued as necessary to keep the consistence with the KL model derivation. Fig.~\ref{werhl_plot}, displays the Wehrl entropy in solid lines (Eq. (\ref{wherlentropy2})) for three different virtualities. It is also shown the two terms in Eq. (\ref{wehrldecomposition}), $S^{\perp}_W$ in dotted lines, and $S_{EE}$ in dashed lines. Along with the line patterns, the color patterns reveals the virtuality values, $Q = 2$ GeV in blue, $Q = 5$ GeV in red and $ Q = 10$ GeV in black. One can see that as the virtuality grows, so the $S_{\text{EE}}$ grows in the small--$x$ regime. The Wehrl entropy on the other hand is weakly dependent on the virtuality. 
In a physical sense, no matter the energy deposition in the proton, its entropy is no longer strongly affected. 
This behavior can be understood from the normalization of the Husimi distribution used in this work, which removes the overall multiplicity factor associated with the gluon density. 
Since the dominant $Q^2$ dependence enters through the total number of resolved gluons, this normalization absorbs most of the scale dependence, leaving mainly the transverse phase--space structure. Then, the mild dependence on $Q^{2}$ is mainly due to the normalization in Eq.~(\ref{normalization}). For the transverse entropy definition $S^{\perp}_W$, as the virtuality grows, the transverse entropy decreases approximately at the same rate.

Notice from the Fig.~\ref{werhl_plot} that $S_W > S_{\text{EE}}$, to give an appropriate interpretation, is need to go back to the respective entropy definitions. The entanglement entropy in KL is purely quantum mechanics and its definition relies on the longitudinal degrees of freedom. On the other hand, the Wehrl entropy is a semiclassical entropy definition that is dependent to the gaussian smooth. It also has four additional degrees of freedom when compared with $S_{\text{EE}}$. Unfortunately, it is no possible to quantify the amount of entropy that rises from this redefinition. But, as in statistical mechanics, these entropies are extensive quantities. From algebra manipulation of $S_W$ in Eq.~(\ref{wehrldecomposition}) we identified a term that is equivalent to $S_{EE}$. Therefore, it is natural to expected the Werhl entropy to be bigger than the entanglement entropy. In this scenario  the first term of Eq.~(\ref{wehrldecomposition}) was defined as the transverse entropy $S^{\perp}_{W}$ . Besides the effects of the gaussian convolution that implies in a semiclassical distribution, it could be a way to quantify the entropy contribution from the transversal degrees os freedom, that are not accessible from the standard $S_{\text{EE}}$. The residual transverse contribution depends on the variables $\mathbf{b}_\perp$ and $\mathbf{k}_\perp$, and therefore encodes transverse phase--space information not contained in the longitudinal entanglement entropy $S_{\text{EE}}(x)$. It can be qualitatively related to the marginal phase--space entropies associated with generalized parton distributions and transverse--momentum distributions. Schematically, one may write $S_W^\perp = S_{\text{sh}}(\text{GPD}) + S_{\text{sh}}(\text{TMD}) - S_{\text{EE}}$, where $S_{\text{sh}}(\text{TMD})$ and $S_{\text{sh}}(\text{GPD})$ denote the Shannon entropies of the transverse momentum and impact--parameter distributions. In this sense, $S_W^\perp$ characterizes the effective transverse phase--space area occupied by the partonic distribution.

The observed hierarchy $S_W > S_{\text{EE}}$ admits a physical interpretation in terms of the amount of information encoded in the underlying degrees of freedom. The Wehrl entropy is a semiclassical quantity defined in phase--space and therefore incorporates both longitudinal and transverse information, providing a description that is closer in spirit to thermodynamic entropy. By contrast, the entanglement entropy in the KL framework is constructed from parton distribution functions, which originate from the parton model and encode only the longitudinal momentum structure of the hadron. In this picture, a fast--moving nucleon is described as a collection of quasi--free partons whose dynamics is fully characterized by PDFs depending on the longitudinal momentum fraction $x$, with no access to transverse spatial or momentum correlations, spin--orbit couplings, or orbital angular momentum \cite{collinear,COLLINS1982445,Martin_1998}. As a consequence, although the KL model successfully reproduces hadronic entropy extracted from multiplicity distributions, its information content is intrinsically limited when compared to phase--space--based entropic measures. The larger magnitude of the Wehrl entropy thus reflects the additional loss of information induced by coarse--graining in transverse phase--space, consistently extending the purely longitudinal entanglement entropy to a more complete semiclassical description.

\begin{figure}[t!]
    \centering
    \includegraphics[scale=0.55]{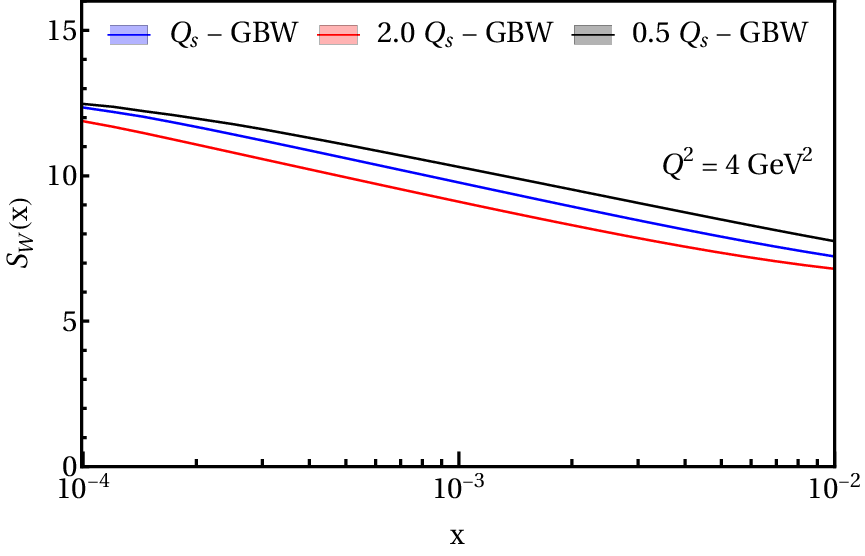}
    \includegraphics[scale=0.55]{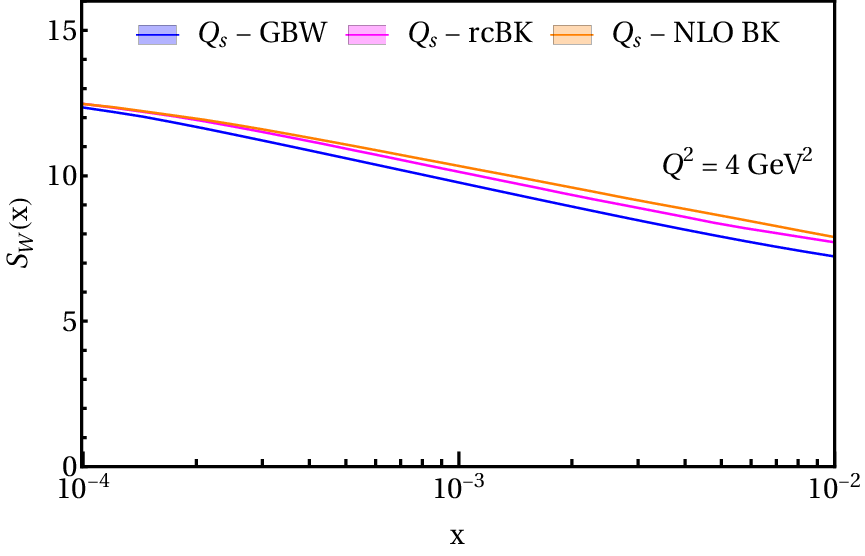}
    \caption{Wehrl Entropy for $Q^{2} = 4$ GeV$^{2}$. Left panel: smearing width $\ell = c/Q_s(x)$, it is shown in blue solid lines $c = 1$, black solid lines $c = 0.5$ and red solid lines $ c = 2.0 $, with $Q_s(x)$ the saturation scale in the GBW model. Right panel: analysis of the smearing width $\ell = 1/Q_s(x)$ with the saturation scale in three different models. Magenta solid line $Q_s(x)$ -- rcBK \cite{rcBK}, blue solid line this $Q_s(x)$ -- GBW and orange solid line $Q_s(x)$ -- NLO BK \cite{NLOBK}. }
    \label{cGBW}
\end{figure}

In addition to the baseline analysis, we explored the dependence of the Wehrl entropy on the Gaussian smearing width and on the choice of the saturation scale. 
To probe the sensitivity to the smearing scale, we considered moderate variations 
of the width following $\ell = c/Q_s(x)$. The left panel of Fig.~\ref{cGBW} shows the 
results for three representative values: $c=0.5$ (black), $c=1.0$ (blue, the same used throughout this work), and $c=2.0$ (red). One observes that the results 
exhibit only a mild dependence on the parameter $c$. It is worth noting that the uncertainty relation $\boldsymbol{\sigma}_{\mathbf{k}_\perp}\boldsymbol{\sigma}_{\mathbf{b}_\perp}=1/2$ is preserved. Changing $c$ effectively redistributes the smearing between position and momentum space: for $c>1$ the distribution becomes narrower in $\mathbf{b}_\perp$ and broader in $\mathbf{k}_\perp$, whereas for $c<1$ the opposite behavior occurs.

A similar analysis was performed using different models for the saturation scale. The right panel of Fig.~\ref{cGBW} shows the results obtained with three choices: the GBW model (blue), the running--coupling BK (rcBK) solution \cite{rcBK} (magenta), and the next--to--leading--order BK (NLO BK) evolution \cite{NLOBK} (orange). A small but visible dependence on the saturation--scale model is observed, as expected. Varying the saturation scale has an effect similar to introducing a constant factor in the smearing width: it preserves the uncertainty relation 
$\boldsymbol{\sigma}_{\mathbf{k}_\perp}\boldsymbol{\sigma}_{\mathbf{b}_\perp}=1/2$ while shifting the 
relative spreading between transverse position and momentum space. The results shown in Fig.~\ref{cGBW} indicate that the qualitative behavior of the Wehrl entropy is robust against reasonable variations of both  the smearing width and the saturation--scale model, suggesting that the main 
physical conclusions are not tied to a specific parameterization.

The entanglement entropy in the Kharzeev–-Levin framework has been extensively studied in the literature and was shown to be in good agreement with experimental multiplicity data in proton-–proton collisions \cite{CMS:2010qvf}. As shown in Fig. \ref{werhl_plot}, the Wehrl entropy is larger than the entanglement entropy, reflecting the additional loss of information induced by coarse--graining in phase--space. For this reason, the Wehrl entropy is not expected to directly reproduce multiplicity data, but instead encodes the degree of decoherence relevant for the emergence of a semi--classical description. Recent work \cite{Rabelo-Soares:2025dfu} has demonstrated that the Wehrl entropy is the appropriate quantity to characterize the initial conditions of hydrodynamic evolution in heavy--ion and small--system collisions. The underlying argument is that, if hydrodynamic evolution in small systems reflects genuine local equilibrium, then the entropy current, rather than the energy-–momentum tensor, provides the physically relevant description of the initial state.

 
\section{SUMMARY AND CONCLUSIONS}
\label{sec:conclusion}

In this work, we investigated the entropy associated with partonic phase--space distributions in QCD, focusing on the Wehrl entropy constructed from Husimi distribution of unpolarized gluons in a unpolarized proton. Starting from the Wigner distribution of gluons, the Husimi distribution was obtained through a Gaussian smearing that ensures positivity (in the analysis performed in this work) and introduce a minimal phase--space resolution consistent with the uncertainty principle. This framework allows for a semiclassical definition of entropy in phase--space, while retaining a clear connection to the underlying the gluonic dynamics. We computed the Wehrl entropy in a phenomenological model for gluon distribution and its dependence on the relevant kinematic variables. By separating the longitudinal and transverse contributions, we identified distinct behaviors associated with each sector. In particular, the longitudinal contribution is directly linked to the normalization of the Husimi distribution and naturally leads to a term that corresponds with the entanglement entropy in KL model. In this sense, the entanglement entropy was defined as a component of the Wehrl entropy. The transverse contribution to the Wehrl entropy encodes information on the spatial and momentum structure of partons in the transverse plane. At small--$x$, the Wehrl entropy decreases with $Q^{2}$, while EE increases, making the Wehrl entropy almost independent of the virtuality, an expected behavior from its intensive nature due to the inclusion of the normalization factor. We further verified that the main conclusions remain stable under reasonable variations of the Gaussian smearing width and the saturation--scale model. The results presented here strongly motivate a more detailed analysis of the Wehrl entropy in the experimental observables. However such a study is beyond the scope of this work, and we plan to explore it in future publications.

\section*{Acknowledgments}

G.R--S. is supported by the CAPES doctoral fellowship 88887.005836/2024--00. R.F. acknowledges support from the Conselho Nacional de Desenvolvimento Cient\'{\i}fico e Tecnol\'ogico (CNPq, Brazil), Grant No. 161770/2022--3. G.S.R. acknowledges support from the Conselho Nacional de Desenvolvimento Cient\'{\i}fico e Tecnol\'ogico (CNPq, Brazil), Grant No. 150338/2026--0. G.T. thanks Bolsa de produtividade CNPQ 305731/2023-8 and FAPESP 2023/06278--2, as well as FAPESP temático 2023/13749--1 for support.

\bibliographystyle{h-physrev}
\bibliography{ref}

\end{document}